# The Hidden Cost of Defaults in Recommender System Evaluation

Hannah Berling*
hannah.berling@gmail.com
University of Gothenburg
Gothenburg, Sweden

Robin Svahn*
robin.svahn@outlook.com
University of Gothenburg
Gothenburg, Sweden

Alan Said
alan@gu.se
University of Gothenburg
Gothenburg, Sweden

## Abstract

Hyperparameter optimization is critical for improving the performance of recommender systems, yet its implementation is often treated as a neutral or secondary concern. In this work, we shift focus from model benchmarking to auditing the behavior of RecBole, a widely used recommendation framework. We show that RecBole's internal defaults, particularly an undocumented early-stopping policy, can prematurely terminate Random Search and Bayesian Optimization. This limits search coverage in ways that are not visible to users. Using six models and two datasets, we compare search strategies and quantify both performance variance and search path instability. Our findings reveal that hidden framework logic can introduce variability comparable to the differences between search strategies. These results highlight the importance of treating frameworks as active components of experimental design and call for more transparent, reproducibility-aware tooling in recommender systems research. We provide actionable recommendations for researchers and developers to mitigate hidden configuration behaviors and improve the transparency of hyperparameter tuning workflows.

*Both authors contributed equally to this research.

## 1 Introduction

Recommender systems research increasingly relies on reusable frameworks to streamline model development, experimentation, and evaluation. Toolkits such as RecBole [19], Elliot [1], and RecPack [13] aim to promote replicability and transparency by providing standardized pipelines for algorithm implementation, dataset access, and evaluation metric computation.

However, despite this shift toward structured tooling, reproducibility in recommender systems remains elusive. Recent analyses [8, 15] highlighted that even papers using open-source frameworks frequently omit crucial configuration details such as hyperparameter tuning procedures, search spaces, and early-stopping criteria that are critical to replicate reported outcomes. Moreover, when such frameworks internally override default behaviors of third-party libraries (e.g., search strategies in `HyperOpt` [5]), even faithful re-runs of published code may yield divergent results.

In this work, we shift the focus from performance benchmarking to *framework-level reproducibility auditing*. Using RecBole as a case study, we examine how built-in hyperparameter search strategies, i.e., Grid Search, Random Search, and Bayesian Optimization (TPE), yield performance differences not solely due to algorithmic merit, but due to undocumented and opaque implementation decisions.

We conduct controlled experiments on six commonly-used recommendation algorithms (EASE, ItemKNN, NeuMF, MultiVAE, RecVAE, and SGL), focusing not on which model or strategy performs best, but on *how* RecBole's internal behavior, such as default early-stopping thresholds and search space interpretation, can affect evaluation outcomes and their how they are reported or understood..

Our findings call attention to:

- how subtle implementation-level defaults shape the conclusions drawn from comparative experiments,
- how randomness and undocumented stopping criteria can obscure reproducibility across repeated runs, and
- the need for recommender systems frameworks to expose and document these behaviors explicitly.

By unpacking these issues, we aim to contribute a methodological reflection on how tooling choices impact scientific claims in the RecSys community. Our goal is not to promote one search strategy over another, but to advocate for more transparent reporting practices and framework accountability.

## 2 Related Work

Reproducibility has emerged as a core concern in recommender systems research. Several studies have shown that even when using public datasets and open-source code, many experimental results are difficult to replicate due to insufficient reporting and non-standardized evaluation practices [6, 8, 14]. A recent systematic review by Bauer et al. [2] further emphasizes the limited diversity in datasets and metrics used across evaluation studies, and highlights the narrow scope of experimentation that dominates current practices. These concerns have led to the introduction of reproducibility tracks at conferences such as RecSys, aiming to encourage transparent and verifiable research.

Hyperparameter tuning plays a critical role in recommendation performance, yet it is often underreported or inconsistently applied. Shehzad and Jannach [15] demonstrated that nearly any model can appear state-of-the-art when compared to poorly tuned baselines. They found that only a minority of papers document their tuning strategy in full, making it difficult to assess whether observed improvements are due to algorithmic advances or better optimization.

Frameworks like RecBole [19], Elliot [1], and RecPack [13] have attempted to address these issues by standardizing experiment pipelines. However, each framework also introduces its own assumptions and behaviors. For example, RecBole integrates with `HyperOpt` [5] and defines default early-stopping criteria internally, such as limiting search continuation after a fixed number of iterations without improvement. These defaults are often not exposed in documentation or configuration files, and can significantly affect tuning outcomes.

The machine learning literature has extensively compared hyperparameter optimization strategies. Grid Search remains the most exhaustive, though computationally expensive [3]. Random Search can be more efficient when only a few hyperparameters influence performance [4]. Bayesian Optimization, particularly via Tree-structured Parzen Estimators (TPE), is known to offer good trade-offs between search efficiency and performance [4]. However, most comparisons assume full control over implementation, which is often not the case when using third-party frameworks with preconfigured logic.

Our work builds on these insights by treating the framework as a first-class object of study. Rather than evaluating which optimization strategy is best, we focus on how built-in configurations, parameter ranges, and stopping criteria affect the reproducibility and transparency of recommendation experiments.

# 3 Method

We examine the behavior of three hyperparameter optimization strategies as implemented in the RecBole framework: Grid Search, Random Search, and Bayesian Optimization via Tree-structured Parzen Estimators. Rather than aiming for state-of-the-art performance, our goal is to assess how internal framework defaults and strategy behaviors impact reproducibility and evaluation outcomes.

## 3.1 Experimental Setup

All experiments were conducted on a Linux Ubuntu 20.04.6 system with an NVIDIA RTX 3090 GPU, using RecBole version 1.2.0 within a Python 3.8 Conda environment. We used the MovieLens-1M [9] and BeerAdvocate [12] datasets, applying standard preprocessing: ratings below 3 were excluded, and users or items with fewer than 10 interactions were removed. For the BeerAdvocate we used the `overall` column for ratings. The source code for our experiments is available on GitHub[1]

[1] https://github.com/alansaid/recsysdefaults

**Table 1: Overview of the selected algorithms, their types, and the number of hyperparameters (HP) and combinations searched.**

| Model | Type | # of HP | # HP Combin. |
|---|---|---|---|
| EASE | GOFAI | 1 | 6 |
| ItemKNN | GOFAI | 2 | 24 |
| NeuMF | Deep Learning | 3 | 60 |
| MultiVAE | Deep Learning | 1 | 5 |
| RecVAE | Deep Learning | 2 | 45 |
| SGL | Deep Learning | 3 | 100 |

**Table 2: The selected hyperparameter ranges for each model used in the search strategies. These values are based on RecBole's documented defaults.**

| Model | Hyperparameter | Values |
|---|---|---|
| EASE | `reg_weight` | 1.0, 10.0, 100.0, 250.0, 500.0, 1000.0 |
| ItemKNN | `k` | 10, 50, 100, 200, 250, 300, 400, 500, 1000, 1500, 2000, 2500 |
| | `shrink` | 0.0, 1.0 |
| NeuMF | `learning_rate` | 0.01, 0.005, 0.001, 0.0005, 0.0001 |
| | `mlp_hidden_size` | [64, 32, 16], [32, 16, 8] |
| | `dropout_prob` | 0.0, 0.1, 0.2, 0.3, 0.4, 0.5 |
| MultiVAE | `learning_rate` | 0.01, 0.005, 0.001, 0.0005, 0.0001 |
| RecVAE | `learning_rate` | 0.01, 0.005, 0.001, 0.0005, 0.0001 |
| | `latent_dimension` | 64, 100, 128, 150, 200, 256, 300, 400, 512 |
| SGL | `ssl_tau` | 0.1, 0.2, 0.5, 1.0 |
| | `drop_ratio` | 0.0, 0.1, 0.2, 0.4, 0.5 |
| | `ssl_weight` | 0.005, 0.05, 0.1, 0.5, 1.0 |

## 3.2 Model and Hyperparameter Selection

To capture a range of tuning behaviors, we selected six RecBole-supported models that vary in complexity and number of hyperparameters. These include two traditional "good old-fashioned AI" (GOFAI) algorithms, EASE [17] and ItemKNN [7], as well as four deep learning-based models: NeuMF [10], MultiVAE [11], RecVAE [16], and SGL [18]. The number of tunable parameters ranged from one (e.g., EASE's regularization weight) to three (e.g., SGL's self-supervision hyperparameters), with search spaces ranging from 5 to 100 combinations, based on the default ranges provided by RecBole documentation.

Table 1 and Table 2 summarize the selected hyperparameter ranges and the corresponding search space sizes used across models.

## 3.3 Search Strategies and Configuration Behavior

All strategies were run using RecBole's `HyperTuning` module. Grid Search evaluated all combinations exhaustively. In contrast, both Random Search and Bayesian Optimization were configured to terminate early if no improvement in the validation metric (nDCG@10) was observed after 10 iterations, a behavior inherited from RecBole's default override of HyperOpt's early-stopping policy. This early-stopping condition is undocumented in the main RecBole documentation but is applied internally unless explicitly overridden.

| Model | EASE | ItemKNN | NeuMF | MultiVAE | RecVAE | SGL |
|---|---|---|---|---|---|---|
| Default | 0.2789 | 0.2345 | 0.1870 | 0.2404 | 0.2625 | 0.2489 |
| Grid | 0.2865 | 0.2391 | 0.2266 | 0.2520 | 0.2633 | 0.2555 |
| Rand 1 | 0.2865 | 0.2390 | 0.2236 | 0.2520 | 0.2625 | 0.2555 |
| Rand 2 | 0.2865 | 0.2382 | 0.2229 | 0.2520 | 0.2614 | 0.2555 |
| Bayes 1 | 0.2865 | 0.2391 | 0.2266 | 0.2520 | 0.2614 | 0.2555 |
| Bayes 2 | 0.2865 | 0.2391 | 0.2199 | 0.2520 | 0.2598 | 0.2554 |

**Table 3: nDCG@10 scores across hyperparameter tuning strategies. Scores for Random and Bayesian Search are shown across two independent runs to illustrate performance variability.**

Each strategy was applied to the same models, using identical search spaces and fixed configuration files. To measure variability in non-deterministic search procedures, we repeated both Random Search and Bayesian Optimization runs twice per model. For each model-strategy pairing, we recorded the number of evaluations performed, the best nDCG@10 score on the validation set, and the configuration responsible for that score.

## 3.4 Evaluation Focus

This study is designed to identify where reproducibility concerns arise not from stochasticity or dataset variance, but from decisions made by the framework itself. Particular attention is given to the influence of early-stopping thresholds, strategy-specific performance variability, and the comparative performance of default configurations relative to tuned ones. This analysis shows how invisible framework logic can bias the reporting and perception of experimental results in recommender systems.

# 4 Results

In this section, we only report the MovieLens-1M results as the results for the BeerAdvocate dataset showed similar results. Our complete suite of results, including those for BeerAdvocate, is available in our code repository.

Across all six models, hyperparameter tuning led to improvements in nDCG@10 compared to the default configurations provided by RecBole (see Table 3). This held true in all but two cases (RecVAE: Random2, Bayes1), with Grid Search yielding the highest scores in most cases, although not by a meaningful margin. These gains confirm that tuning matters, but our focus lies in how these gains are reached, and how sensitive they are to RecBole's internal configuration logic.

For models with small search spaces such as EASE (6 combinations) and MultiVAE (5 combinations), all three strategies found identical optimal configurations. This is unsurprising, as both Random Search and Bayesian Optimization effectively performed near-exhaustive searches due to RecBole's early-stopping default of 10 non-improving iterations. In these settings, differences in runtime were negligible, and performance converged quickly.

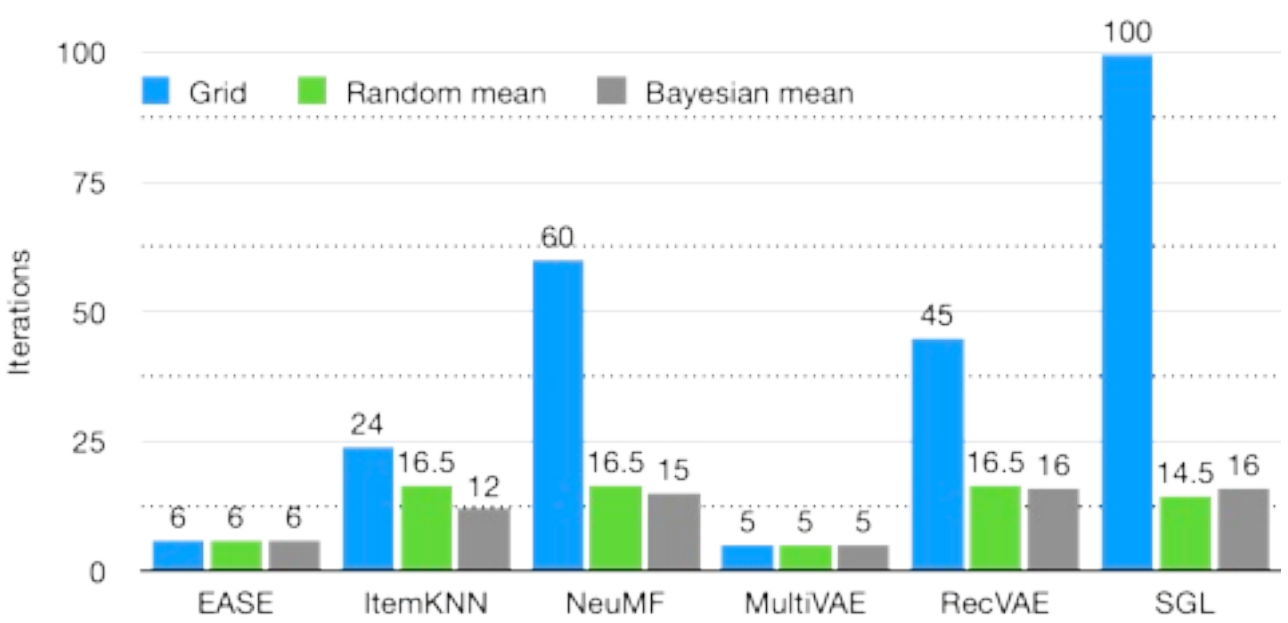

**Figure 1: Number of iterations required for each search strategy to terminate per model. Both Random Search and Bayesian Optimization use RecBole's undocumented `early_stop = 10` policy, causing most runs to terminate after 10–12 iterations, even for models with large search spaces (e.g., NeuMF: 60, SGL: 100). This early termination prevents full exploration of the search space and can mislead users about the thoroughness of tuning. (Means shown for repeated runs of stochastic methods.)**

However, for models with larger search spaces such as RecVAE (45 combinations) or NeuMF (60 combinations), search behavior diverged more noticeably (see Fig. 1). Surprisingly, the model with the highest number of parameters, SGL (100 combinations), appeared unaffected by the search strategy. Although Grid Search consistently found the best configuration, it did so at a much higher computational cost. To quantify this, we recorded the elapsed time for each Grid Search run, rounded to the nearest minute. The simpler models completed quickly: EASE in 1 minute, ItemKNN in 12 minutes, and MultiVAE in 19 minutes. In contrast, deeper models required significantly longer: NeuMF ran for 199 minutes, RecVAE for 260 minutes, and SGL for 985 minutes. This demonstrates how runtime grows with both model complexity and search space size. A closer look at the actual search paths for these strategies, visualized in Fig. 2, shows that Bayesian Optimization tends to concentrate its sampling in performance-promising regions, while Random Search, as expected, explores more broadly and erratically.

An important finding is the extent to which RecBole's undocumented defaults shape outcomes. Both Random Search and Bayesian Optimization use an early-stopping policy set to 10 by default, meaning they terminate once 10 evaluated configurations fail to improve validation performance. This stopping criterion, which is not surfaced in configuration files or logs, introduces an implicit assumption that can directly affect replicability and fairness in model comparisons. In our experiments, this policy was triggered in all runs on models with more than 10 configurations, including NeuMF (60 combinations), RecVAE (45), and SGL (100), typically halting the search after 10–12 iterations. As a result, large portions of the search space were never explored, despite the illusion of a complete or thorough search. Without careful inspection of logs or source code, users may mistakenly believe that the full search budget was exhausted, when in fact it was silently truncated.

We also note that performance differences between strategies are often smaller than the differences between repeated runs of the same strategy. For instance, the average difference in best

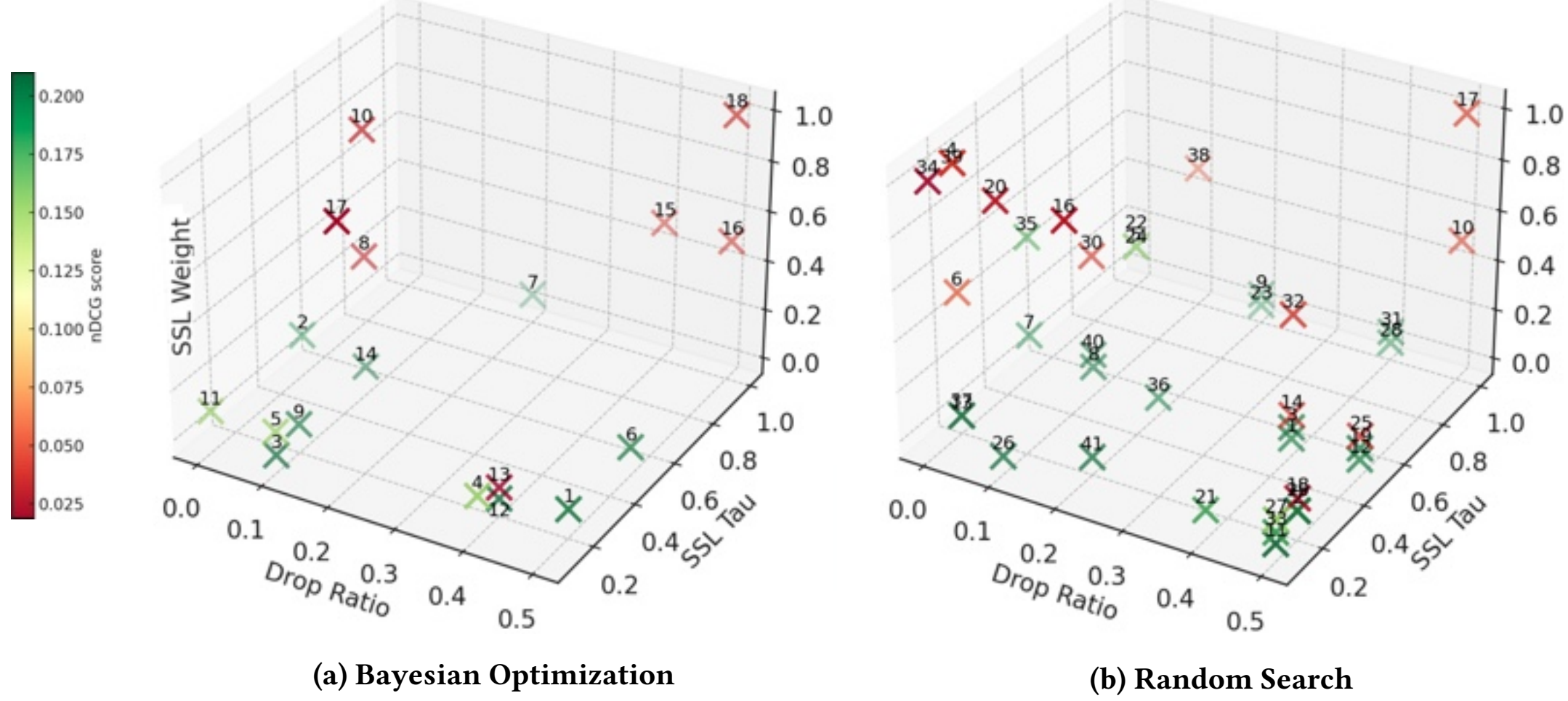

**(a) Bayesian Optimization** **(b) Random Search**

**Figure 2: Scatter plot of search space traversal for SGL. The number on each data point refers to the iteration, the color indicates the nDCG@10 score given the combination of hyperparameters (drop_ratio, ssl_tau and ssl_weight)**

nDCG@10 between the two runs of Bayesian Optimization on RecVAE was comparable to the difference between Bayesian Optimization and Random Search. This variance further underscores the need to report search settings and framework behaviors in detail, especially when evaluating new models against tuned baselines.

Due to computational constraints, each stochastic strategy was run twice per model. While this limits statistical testing, we observed notable variation in some cases, for example, NeuMF's best nDCG@10 differed by 0.0067 between two Bayesian Optimization runs, and by the same amount between Bayesian Optimization and Grid Search. While small in absolute terms, these differences occurred despite identical search spaces and suggest that RecBole's early-stopping behavior can lead to inconsistent outcomes depending on search dynamics and randomness. This reinforces the need for repeated trials and careful documentation of tuning behavior, especially when evaluating stochastic search methods.

In summary, while all strategies improved over defaults, their efficiency and determinism varied. RecBole's internal behaviors, particularly early stopping and opaque tuning logic, have a measurable impact on what results are obtained and how they are interpreted. These effects are not always visible to the user but are crucial for reproducibility.

## 5 Discussion

While all search strategies evaluated in this study improved over the default hyperparameter settings in RecBole, the performance differences between them were often modest. However, the differences in how each strategy explores the hyperparameter space, and the extent to which they are influenced by RecBole's internal configuration logic, were substantial.

The most prominent example of this is the early-stopping behavior applied to both Random Search and Bayesian Optimization. By default, RecBole halts these strategies after 10 iterations without improvement in the validation metric. This behavior is not declared in configuration files, nor is it clearly documented, yet it affects how long each strategy runs and how thoroughly the search space is explored. As shown in Figure 1, strategies often terminated before the total number of available configurations had been tested. In models with many parameters, like NeuMF, this likely means that many candidate configurations were never evaluated.

This internal behavior poses a challenge for reproducibility. Since most papers using RecBole do not report implementation-specific search settings, the same experiment may be irreproducible without access to the framework version and tuning logs. Even when results are numerically similar across strategies or repeated runs, as shown in Table 3, the underlying search paths differ significantly. These differences are rarely visible to the researcher, let alone to readers of a published paper.

A summary of practical recommendations for researchers is provided in the box below.

**Guidance for Researchers**

To avoid misleading or irreproducible tuning outcomes when using RecBole or similar frameworks:

- **Check hidden defaults.** Some behaviors (e.g., early stopping) are undocumented and must be verified in source code.
- **Override early stopping.** Set `early_stop=None` or raise the threshold to ensure full search.
- **Log search details.** Save all configurations, seeds, scores, and stopping conditions.
- **Use multiple seeds.** Repeat stochastic searches to gauge variability.
- **Report all tuning settings.** Include strategy type, search budget, early stopping rule, and framework version in your paper.

Our findings also suggest that variability between repeated runs of stochastic strategies can rival the difference between those strategies themselves. This reinforces the idea that benchmark results should be interpreted in the context of search dynamics, not just validation scores. A single run may not be representative, particularly when tuning is sensitive to randomness and prematurely terminated by internal heuristics.

Although we did not conduct formal significance testing due to the limited number of runs per strategy, the observed variability between repeated trials suggests that performance differences between strategies may not always be meaningful. With only two runs per strategy, we cannot reliably distinguish between algorithmic differences and randomness introduced by internal framework behavior. This limitation reinforces our broader argument: reproducibility in recommender systems requires more than just access to code – it depends on transparent tuning procedures, sufficient repetitions, and visibility into framework defaults.

More broadly, our results support the position that recommender systems research should treat frameworks like RecBole not just as neutral tooling, but as active participants in experimental outcomes. Internal defaults, undocumented behaviors, and implicit design choices have the potential to shape conclusions, especially in performance-sensitive comparisons between algorithms.

To improve reproducibility, we encourage researchers to document not just hyperparameter ranges and metrics, but also the full configuration of the search strategy, including stopping criteria, evaluation budget, and the specific framework version used. Framework developers, in turn, should expose these defaults explicitly and encourage the use of configuration files that make them transparent and reproducible by design.

This study has several limitations. First, our experiments were limited to two datasets, MovieLens-1M and BeerAdvocate, and a fixed set of six models. While this setup allows for a controlled analysis of framework behavior, it does not capture the full range of tuning challenges that might arise with other datasets or model architectures. Second, we focused exclusively on search strategies natively supported by RecBole, and did not evaluate the effect of user-defined search spaces, custom stopping rules, or hybrid optimization approaches. Third, while we concentrated on RecBole as a case study, we believe that similar undocumented defaults and reproducibility risks may exist in other frameworks such as Elliot or RecPack, especially as they evolve. Finally, our analysis is specific to RecBole version 1.2.0, and future versions may expose or alter some of the undocumented behaviors we highlight. These constraints suggest that our findings should be seen as indicative rather than exhaustive, but they underscore the broader point that reproducibility in recommender systems depends critically on surfacing and documenting configuration-level decisions.

## 6 Conclusions

This paper presented a reproducibility-focused audit of hyperparameter search behavior in RecBole. While all tested strategies (Grid Search, Random Search, and Bayesian Optimization) led to performance gains over default configurations, their outcomes were shaped not only by their inherent logic, but also by RecBole's internal defaults and undocumented configuration policies.

In particular, we showed that RecBole's built-in early-stopping logic for stochastic strategies, which is not externally visible or configurable through standard YAML files, significantly limits the number of configurations evaluated. This has important implications for both fairness in model comparisons and the reproducibility of published results. Our findings also revealed that performance variability across repeated runs of the same strategy can match or exceed the differences between strategies themselves.

These observations point to a broader challenge in recommender systems research. Frameworks such as RecBole are essential tools for standardization and scalability, but their internal behavior must be treated as part of the experimental design. Transparent reporting of tuning strategies, early-stopping criteria, and framework versions is essential for reproducible, trustworthy comparisons. While our audit focused on RecBole, we believe similar issues likely exist in other toolkits, particularly as they grow more complex.

We provided practical recommendations for both researchers and developers to mitigate hidden behaviors and document experimental pipelines more rigorously. Only by treating tooling as part of the methodological stack can we ensure that future improvements in recommender systems are both meaningful and reproducible.

## 7 Recommendations for Developers

To improve transparency and reproducibility, frameworks like RecBole should make internal behaviors such as early stopping and search strategy defaults clearly visible and configurable. Default parameters should appear in configuration files or logs rather than being hidden in the source code. Search logs should include enough detail, including evaluated configurations, random seeds, and stopping conditions, to allow others to reproduce tuning runs exactly. Maintaining versioned documentation that tracks changes in tuning behavior would also help researchers interpret results across different versions. Frameworks are not passive tools but active parts of the experimental setup, and should be treated as such.